# Laboratory Characterisation bench for high precision astrometry

Fabrice Pancher*[a], Sebastien Soler[a], Fabien Malbet[a], Manon Lizzana[a,b,c], Pierre Kern[a], Thierry Lepine[d], Alain Leger[e]

[a] Université Grenoble Alpes, CNRS, IPAG, Grenoble, France; [b] Centre National d'Études Spatiales, Toulouse, France; [c] Pyxalis, Moirans, France ; [d] Université Jean Monnet - Saint-Etienne, IOGS, CNRS, Laboratoire Hubert Curien, Saint-Etienne, France ; [e] Université Paris-Saclay, CNRS, CNES, IAS, Gif-sur-Yvette, France

**ABSTRACT**

High precision differential astrometry assesses the positions, distances, and motions of celestial objects in relation to the stars. The focal plane of such space telescope must be calibrated with a precision down to the level of 1e-5 pixel in order to be able to detect Earth-like planets in the close vicinity of the Sun. The presented characterization bench is designed to improve the technology readiness level for the following key points: calibration of new detectors with a high number of pixels and correcting the field distortion using stars in the field of view. The first aim of the project concentrates on the characterization of a 46 megapixels sensor from PYXALIS, to assess its typical parameters using an integrating sphere. The next objective intends to map the intra and extra pixel quantum yield of the detector with a precision of 1e-5 pixels and investigate the evolution of the pixel geometry in response to environment fluctuations. To conduct these tests, an optical bench is designed with an LCD screen and a doublet, used as a source that allows directing light to specific groups of pixels. Interferometric calibration of the detector pixel centroid position will be achieved using fibers that illuminate the detector with Young's fringes. To characterize the distortion of the detector, a diaphragm will produce adjustable optical aberrations to be corrected and therefore change the source sensor positional relationship. The final step involves the simulation of a star's field, which will be imaged on the detector to assess optical quality.

## 1. INTRODUCTION

High-precision differential astrometry measures the positions, distances, and motions of celestial objects relative to the stars [1]. One of the primary scientific objectives, given its demanding nature, is the discovery of Earth-like planets orbiting in the close vicinity of Sun-like stars. Another key goal is to explore the nature of dark matter [1]. To accomplish these objectives, the focal plane of a such space telescope must be calibrated with extreme precision, down to 1e-5 pixels [1, 2, 3, 4]. This level of precision is essential to detect the reflex motion of a star, measured at 0.3 µas, which corresponds to 5e-5 pixels caused by an Earth-like planet orbiting a Sun-like star located 10 pc from us. Two critical challenges arise from these scientific objectives: calibrating new detectors with extremely high pixel counts and correcting field distortion using stars within the field of view. The aim of this paper is to present the experimental setup selected for the characterization of a 46-megapixel imaging sensor as part of the THEIA astrometry project.

The initial phase of the project centers on characterizing the GIGAPYX4600 sensor from PYXALIS [5], to assess its key parameters. The second phase involves the interferometric calibration of the detector, aimed at determining the positions of its pixel's centroids using Young's fringes on the detector. In the third phase, the project will validate the field distortion calibration method using simulated stars created on an LCD screen.

*fabrice.pancher@univ-grenoble-alpes.fr; phone +33 4 76 14 37 04; https://ipag.osug.fr

## 2. STANDARD DETECTOR CALIBRATION

For this initial phase, we used an integrating sphere from the Labsphere HeliosPlus family *USLR-D12F-NDNN-P*, which features a spatial luminance uniformity over the exit port of ±1% and a maximum luminance of 50,000 cd/m². The integrating sphere is equipped with three independently switchable halogen lamps and an adjustable diaphragm that allows to vary the output flux. Control of the sphere is achieved through proprietary software. However, a control library has been developed to interface with the sphere remotely via TCP/IP, enabling integration of this control into Python scripts connected to other devices. Due to hysteresis, the diaphragm control is not very precise. Nevertheless, Labsphere offers a *Hunt and Seek* feature that regulates the diaphragm position based on the light measurement from the sensor integrated into the sphere, facilitating easily reproducible luminosity levels for various experiments.

The Gigapyx sensor is integrated into Genepyx, an electronic enclosure provided by Pyxalis. A lens mount attached to the Genepyx enables access to the sensor. Through the Genepyx, users can configure the sensor and initiate image acquisition via a Thunderbolt connection to a computer. Pyxalis also provides a software development kit (SDK) that facilitates interfacing with the sensor. A custom Python library was developed that enables to define acquisition parameters, particularly focusing on the exposure time and sensor gain for the planned measurement series. Additionally, several internal parameters of the sensor, such as analog and digital gains and offsets, are accessible.

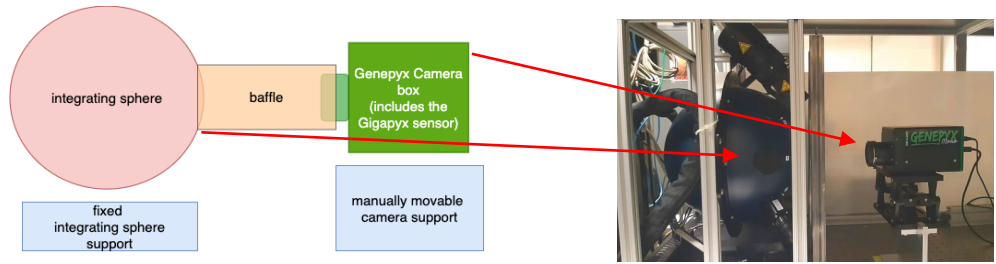

Figure 1. standard detector calibration bench

To characterize the standard parameters of the sensor, the Genepyx entrance is positioned at the output of the integrating sphere. Test automation is facilitated through scripts that interface with both the sensor and the sphere, enabling their configuration, image acquisition, and storage of the collected images in FITS format.

It has been determined [6] that a series of 600 images is required to accurately evaluate the characteristics of the sensor under examination, which amounts to approximately 60 GB per dataset. To extract the sensor characteristics from this data, the FITS images cubes are sent to a remote file server, which is accessible by a high-memory computing server capable of handling such large files. Calibration results obtained through this setup are described in reference [1].

## 3. INTERFEROMETRIC CALIBRATION OF THE PIXELS CENTROID POSITION

In an actual sensor, at a scale smaller than 1/1000 of a pixel, the pixels are not perfectly aligned. To characterize this misalignment, an interferometric calibration of the detector's pixels centroid position is considered. This calibration is performed using optical fibers as the illumination method. The interference of two beams creates Young's fringes [7], which illuminate the detector.

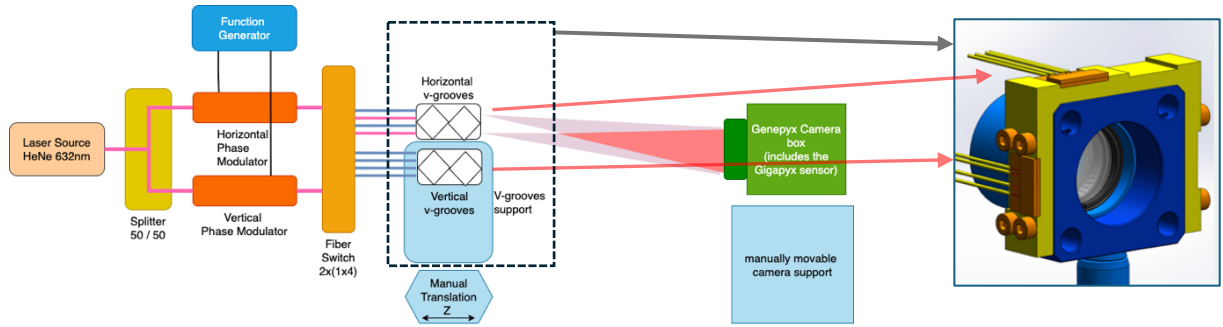

Figure 2. interferometric detector calibration bench

The input signal from the sources is a HeNe laser emitting at 632 nm, which is then split by a 50-50 splitter. Each output from the splitter is modulated by a phase modulator that makes the fringes scroll over each pixel. These modulators are specifically optimized for this wavelength range. The modulator outputs are connected to a 2x(1x4) fiber switch, allowing the signals to be routed to any combination of two groups of four outputs. Four of these outputs are connected to a horizontal 4-fiber V-groove to generate Young's fringes in one direction, while the other four are connected to a vertical 4-fiber V-groove to generate fringes in the perpendicular direction. The V-grooves are mounted on a square lens holder, positioned at the center of the bench used for distortion characterization in the setup.

## 4. OPTICAL DISTORSION CALIBRATION

By simulations, a polynomial model can be applied to calibrate the telescope's distortion with an accuracy better than 5e-6 pixels [6]. It has been demonstrated that reference stars within the telescope's field of view can serve as metrological sources for determining the distortion function [8].

The objective of this setup is to replicate a similar environment by generating pseudo-stars on an LCD screen, introducing distortion with a diaphragm, capturing the images on the detector and finally applying a polynomial model to recover the undistorted image.

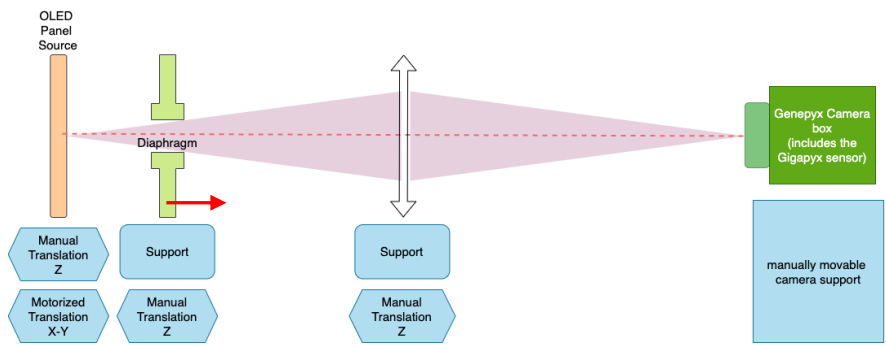

Figure 3. optical distortion calibration bench

To carry out these tests, we set up an optical bench utilizing the Microoled MDP04B, a 5-megapixel OLED monochrome screen with a 0.61-inch size, a pixel pitch of 4.7x4.7 µm, and a luminance of up to 1000 cd/m². A custom Python library was developed to display custom patterns on the screen, allowing the light to be directed to a selected area of pixels on the detector. A 200 mm focal length achromatic doublet is placed in the center of the optical path to ensure that a single pixel from the source reaches at least one pixel of the detector. To characterize the detector's distortion, a diaphragm with variable position along the optical axis is used to introduce adjustable optical aberrations that can be corrected, thereby altering the positional relationship between the source and the sensor. To cover the entire surface of the detector, the OLED source is mounted on three-axis stages, motorized for tip-tilt adjustments and manually controlled for defocus.

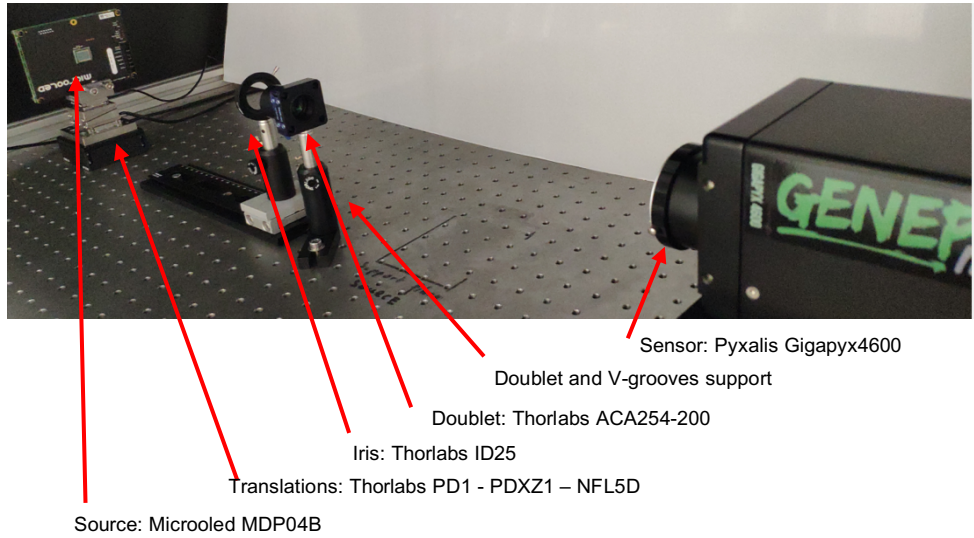

Figure 4. laboratory optical distortion calibration bench

## 5. CONTROL ENVIRONNEMENT

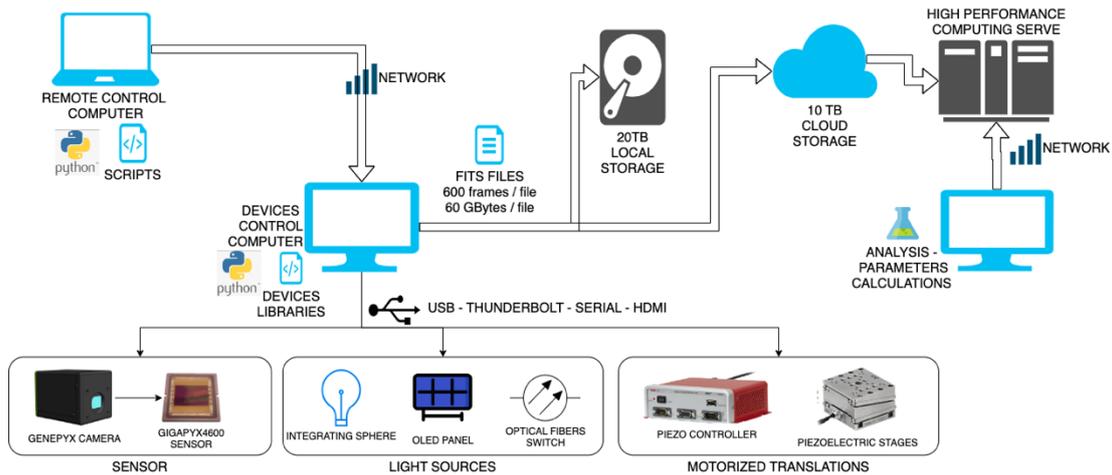

Figure 5. hardware and software control environment

Custom Python libraries have been developed to control all the equipment that can be operated remotely. It includes configuring the Gigapix4600 sensor parameters and capturing images, controlling the integrating sphere elements to generate a reproducible light flux, controlling the fiber switch to select combinations of fiber outputs directed toward the V-grooves, and operating the piezo controllers for the tip-tilt stages that facilitate translations of the OLED screen.

Users can use these libraries to integrate them into their own scripts for remote control of the equipment, enabling automated and reproducible measurement sequences.

The images captured by the sensor are recorded as FITS image cubes, which can be stored locally on the computer connected to the devices or on a remote file server. This server is accessible via a high-performance computing server, allowing the processing of large generated files, typically around 60 GB.

## 6. CONCLUSION

We presented three optical benches aimed at characterizing the GIGAPYX4600 sensor from Pyxalis, which is part of a family of larger detectors that will be developed soon. The first bench, using an integrating sphere, allowed us to characterize the standard parameters of the detector to verify consistency with manufacturer data, and to qualify our setup. The second bench was designed to launch interference Young's fringes onto the sensor, both in one direction and perpendicularly, to precisely locate the centroid of each pixel. Finally, the third bench enables us to reproduce a star field using an LCD screen, where we simulate distortion by translating an iris along the optical axis. This setup will allow us to calculate polynomials to correct the distortion, validating this approach for correcting distortions in a real environment using a field of reference stars. These benches could be used with new larger detectors.

## ACKNOWLEDGEMENTS

The authors would like to thank the researchers and engineers who are not co-authors of this paper but who have taken part and have brought their contribution in the concretization of this project. Concerning the funding of our work, we would like to acknowledge support by the LabEx FOCUS ANR-11-LABX-0013, the CNES agency and PYXALIS.

## REFERENCES

[1]  Malbet, F., Boehm, C., Krone-Martins, A., et al. 2021, Experimental Astronomy, 51, 845
[2]  Shao, M., Zhai, C., Nemati, B., et al. 2023, PASP, 135, 074502
[3]  Ji, J.-H., Li, H.-T., Zhang, J.-B., et al. 2022, Research in Astronomy and Astrophysics, 22, 07200
[4]  Nemati, B., Shao, M., Gonzalez, G., et al. 2020, in Society of Photo-Optical Instrumentation Engineers (SPIE) Conference Series, Vol. 11443, Space Telescopes and Instrumentation 2020: Optical, Infrared, and Millimeter Wave, ed. M. Lystrup & M. D. Perrin, 114430O
[5]  Michelot J., Douix ., Mancini J-B., Guillon M., Melendez K., Ravinet C., Jouans M., Estaves G., Marec R., Demiguel S., Materne A., Virmontois C., "Gigapyx sensor performance in space environments", ICSO (2024)
[6]  Lizzana M., Kern P., Leger A., Lepine T., Malbet F., Pancher F., "Experimental tests of the calibration of high precision differential astrometry for exoplanets and dark matter," ICSO (2024)
[7]  Crouzier, A. et al. *A&A*. 2016, 595, A108
[8]  Malbet, F., Labadie, L., Sozzetti, A., et al. 2022, in Society of Photo-Optical Instrumentation Engineers (SPIE) Conference Series, Vol. 12180, Space Telescopes and Instrumentation 2022: Optical, Infrared, and Millimeter Wave, ed. L. E. Coyle, S. Matsuura, & M. D. Perrin, 121801F